\documentstyle[12pt]{article}
\nopagebreak

\def\oneh{{\textstyle {1\over 2}}}

\def\eq{\begin{equation}}
\def\ee{\end{equation}}
\def\eqa{\begin{eqnarray}}
\def\eea{\end{eqnarray}}

\def\bra#1{\mbox{$\langle #1\vert $}}
\def\ket#1{\mbox{$\vert #1\rangle$}}

\def\plusqrt#1{\hbox{\raise 3pt\hbox{$^+$}\kern-5pt\hbox{$\sqrt{#1}$}}}

\begin{document}

%%%%%%%%%%%%%%%%%%%%%%%%%%%%%   Title   %%%%%%%%%%%%%%%%%%%%%%%%%%%%%%%%%%%

\centerline{\Large{\bf Semi-relativistic charge-current density operator}}

\vskip 1.0cm

\centerline{\large{S.~Boffi, F.~Capuzzi, P.~Demetriou and M.~Radici}}

\vskip 1.0cm

\centerline{\small{Dipartimento di Fisica Nucleare e Teorica,
Universit\`a di Pavia, and}}

\centerline{\small{Istituto Nazionale di Fisica Nucleare, 
Sezione di Pavia, Pavia, Italy}}

\vskip 1.5cm

%%%%%%%%%%%%%%%%%%%%%%%%%%%%%%% Abstract %%%%%%%%%%%%%%%%%%%%%%%%%%%%%%%%%%

\begin{abstract}

\noindent 
The charge-current density and two-photon operators consistent with a
single-particle semi-relativistic Hamiltonian are derived within a suitable
functional derivative formalism which preserves gauge invariance. An
application to electron scattering is presented and results are compared with a
fully relativistic case and the non-relativistic cases corrected through fourth
order in $M^{-1}$.

\end{abstract}

\bigskip

%PACS numbers: 

%{\sl Keywords\/}: charge-current density operator, 
%semi-relativistic Hamiltonian, gauge invariance

\vskip 1.5cm

\clearpage

%%%%%%%%%%%%%%%%%%%%%%%%%%%%%%%   Text   %%%%%%%%%%%%%%%%%%%%%%%%%%%%%%%%%%

\begin{section}{Introduction}

The structure and dynamics of strongly interacting systems, such as nuclei and
nucleons, is best studied using an electromagnetic probe. By treating the
electromagnetic interaction within quantum electrodynamics theoretical
uncertainties are minimized and the relevant information is simply related to
charge- and current-density distributions (for a recent review,
see~\cite{[book]}). However, ambiguities may arise due to an inconsistent
treatment of the different theoretical ingredients belonging to the complex
description of the underlying many-body problem. 

In particular, in connection with relativity one meets two orders of problems.
On the one hand, the elementary operator is well known. To be specific, let us
start considering a free spin-$\oneh$ particle with mass $M$. In terms of
normalized spinors, the matrix element of the four-current describing its
transition from an initial state  $\ket{{\vec p},s}$ (with spinor $u_i$ and
total energy $E_{\vec p} = \sqrt{\vert {\vec p}\vert^2 +  M^2}$) to a final
state  $\ket{{\vec p}',s'}$ (with spinor $u_f$ and total energy $E_{{\vec p}'} =
\sqrt{\vert {\vec p}'\vert^2 +  M^2}$) is written as
\eq
\bra{{\vec p}',s'}j^\mu\ket{{\vec p},s} = {M\over\sqrt{E_{\vec p}E_{{\vec p}'}}}
\overline u_{\rm f}  \left [ \gamma^\mu F_1(Q^2) +  {\rm i}\sigma^{\mu\nu} 
q_\nu {\kappa\over 2M} F_2(Q^2)\right ]u_{\rm i}
(2\pi)^{-3}\,{\rm e}^{-{\rm i}{\vec q}\cdot{\vec x}} , 
\label{eq:cc2}
\ee
where $Q^2 = -q_\nu q^\nu = {\vec q}^2 - \omega^2$ is defined in terms of the
energy $\omega$ and momentum ${\vec q}$ of the absorbed photon (${\vec q} =
{\vec p}' - {\vec p}$); $F_1$ and $F_2$ are the Dirac and Pauli form factors,
respectively, and $\kappa$ is set equal to the anomalous part of the magnetic
moment of the particle.  Alternatively, making use of the Gordon decomposition
one writes

\eq
\bra{{\vec p}',s'}j^\mu\ket{{\vec p},s} 
= {M\over\sqrt{ E_{\vec p} E_{{\vec p}'} }}\, \overline u_{\rm f} 
[(p^\mu  +  {p'}^\mu)\Gamma_1(Q^2) +  \gamma^\mu\Gamma_3(Q^2)]
u_{\rm i} (2\pi)^{-3}\,{\rm e}^{-{\rm i}{\vec q}\cdot{\vec x}},
\label{eq:cc1}
\ee
where
\eq
F_1(Q^2) = 2M \Gamma_1(Q^2) +  \Gamma_3(Q^2), 
\qquad  \kappa F_2(Q^2) = - 2M \Gamma_1(Q^2).
\ee

The current of a composite system is often calculated in the impulse
approximation, i.e. under the assumption that it is given by the
sum of the currents of the individual constituents, treated as free
particles. Under this assumption, the system current results from the
sum of terms like one of the above currents. However, the use of a
relativistic expression for the current inside matrix elements requires that
the corresponding states are also treated relativistically. This is not always
possible in many-body theories of strongly interacting systems where most
information on the structure of nuclei and baryons comes from solving the
non-relativistic Schr\"odinger equation. Therefore, one is faced with the
problem of a non-relativistic reduction of the charge-current density operator.

There are two ways to obtain a non-relativistic approximation including
lowest-order relativistic corrections, when constructing an effective 
Hamiltonian which is to be used in non-relativistic perturbation
theory~\cite{[Fearing]}. In the first, one evaluates the matrix element of the
potential operator between positive energy solutions of the free Dirac
equation and then performs a two-component reduction (direct Pauli reduction).
Low-energy theorems and gauge invariance are satisfied at the price, e.g., that
the electromagnetic interaction of quarks contains a non additive (two-body)
term and the effects of the quark binding potential~\cite{[Close]}.

The second way is the Foldy--Wouthuysen (FW)  method~\cite{[FW]}. It is shown
that the two methods are equivalent to first order in perturbation theory, but
differ to higher order. Moreover, only the FW effective $S$-matrix reproduces
the full relativistic $S$-matrix~\cite{[Fearing]}.

The FW method looks for a canonical transformation which decouples
the Dirac equation into two two-component equations, one reducing to the 
Pauli description in the non-relativistic limit, the other describing 
negative-energy states. Relativistic corrections can be classified according
to the expansion in powers of a suitable parameter. A natural scale can be
identified with $Q/2M$. The formal series expression of the transformed
Hamiltonian must be truncated at some order which will correspond to a
non-relativistic expansion of the transformed Hamiltonian in a power series in
$M^{-1}$. This was done through second order by McVoy and Van
Hove~\cite{[McVoy]} and through fourth order by Giusti and Pacati~\cite{[GP]}.
However, in electromagnetic interactions with nuclei a new scale
emerges essentially related to the large anomalous magnetic moment $\kappa$ of
the nucleon. Therefore, $\kappa Q/2M$ must also be considered as a suitable
expansion parameter when extending the expansion to larger
momenta~\cite{[Bill]}. 

The FW method has been successfully applied to inclusive and semi-inclusive
scattering of electron by nuclei~\cite{[book]}, but may easily break down when
increasing the energy $\omega$ and momentum ${\vec q}$ transferred by the
photon. 

On the other hand, the non-relativistic description of the many-body system
itself may become too crude an approximation when one is exploring extreme
conditions in nuclei or simply considers that in quark models relativistic
effects are most important because the mean velocity of a constituent
quark in the nucleon is comparable with the velocity of light. The
relativistic dynamics of interacting particles is nontrivial and two different
approaches have been used to investigate relativistic effects in binding
energies. For nuclei the theory is patterned after quantum electrodynamics and
is relativistically covariant (for a review see \cite{[Serot]}). In this
approach the antinucleon degrees of freedom play an important role. However,
there is no clear evidence of a simple coupling of $N{\overline N}$ states to
mesons and photons. In the other approach, starting from a relativistically
covariant description of a system of interacting particles where the
interactions are direct rather than mediated through a
field~\cite{[BT],[Foldy],[KF],[Friar]}, not manifestly covariant but
relativistic Hamiltonians have been proposed to study both
hadrons~\cite{[semirel],[Godfrey],[Isgur],[Warns],[Plessas]} and
nuclei~\cite{[Glockle],[Carlson],[Forest],[Forestdue]}. In all cases the
kinetic energy operator was taken as
\eq T = \plusqrt{{\vec p}^2 + M^2}.
\label{eq:kinetic}
\ee 
By this choice one excludes negative energy states {\sl ab initio\/} and
simply solves a Schr\"odinger equation for bound states. Relativized models
have been developed accordingly and also applied. e.g., to investigate
electroexcitation of baryon 
resonances~\cite{[Capstick],[Capsticktwo],[Warnstwo],[Warnsthree]}. 

The problem of finding a charge-current density operator consistent with a
single-particle semi-relativistic Hamiltonian, i.e. a Hamiltonian consisting
of a kinetic energy given by eq. (\ref{eq:kinetic}) and of an external, energy
and momentum independent potential, is addressed in sect. 3. The
derivation makes use of the functional derivative formalism developed in sect.
2 to ensure gauge invariance (see also \cite{[Aren]}).

In this paper we will not consider the fact that hadrons are bound in
the  many-body system and a proper modification of the current is
necessary in order to account for off-shell effects that can only be specified
when a sufficient knowledge of the internal dynamics is available. A variety
of prescriptions have been proposed to solve this problem in specific
cases (see, e.g., \cite{[Frullo],[deForest],[Naus],[Nauslet]}). Ambiguities
due to different prescriptions are comparable and in some cases much larger
than relativistic corrections. In the application to electron scattering
presented in sect. 4 we shall therefore omit off-shell corrections and compare
results with different transition operators under the same kinematic
conditions. Concluding remarks are given in sect. 5.

\end{section}

\begin{section}{Gauge invariance and charge-current density}

Let us consider a point particle with charge $e$, momentum $\vec p$ and 
Hamiltonian $H_0(=p^0)$. According to the usual prescription, the interaction
Hamiltonian with the electromagnetic field is obtained by the minimal
substitution \eq
p^\mu \to p^\mu - e A^\mu, \qquad (p^\mu = {\rm i}\partial^\mu),
\ee
where $A \equiv A^\mu =(A^0, A^i)$ is the electromagnetic field operator whose
time dependence is understood. Greek indices run over 0, 1, 2, 3 and Latin ones
over 1, 2, 3, with a spatially negative metric tensor such that, e. g.,
$x^i=-x_i$, $x^0=x_0=t$ ($\hbar=c=1$).

The resulting Hamiltonian $H(A)$ is gauge invariant, i.e. the physical
properties of the system remain unchanged under a gauge transformation of the
field,
\eq
A^\mu \to {A'}^\mu = A^\mu - \partial^\mu\lambda ,
\ee
where $\lambda({\vec x},t)$ is an arbitrary function satisfying the D'Alembert
equation. The requirement that the Schr\"odinger equation is conserved in
form under the transformation of the Hamiltonian
\eq
H(A) \to H'(A) = H(A') ,
\ee
when accompanied by the local phase transformation of the wavefunction
\eq
\Psi({\vec x},t) \to \Psi'({\vec x},t) 
= {\rm e}^{{\rm i}e\lambda({\vec x},t)} \Psi({\vec x},t) ,
\ee
leads to the identity in $\lambda(t)$ and $\dot{\lambda}(t)$,
\eq
H(A') + e \dot{\lambda}(t) = 
{\rm e}^{{\rm i}e\lambda(t)} H(A) {\rm e}^{-{\rm i}e\lambda(t)} ,
\label{eq:lambda}
\ee
where $\lambda(t)$ and $\dot{\lambda}(t)$ are operators multiplying the
wavefunction by $\lambda({\vec x},t)$ and $\partial_t\lambda({\vec x},t)$, 
respectively.

The interaction Hamiltonian with the electromagnetic field,
\eq
H_{\rm em}(A) \equiv H(A) - H_0 ,
\ee
can usefully be expanded in a Taylor series of functional derivatives. Within
the first quantization formalism it is the expectation value of an operator
$O(A)$ given as a functional of the four components $A^\mu({\vec x},t)$ to
be considered as functions of ${\vec x}$ depending on the parameter $t$. The
functional derivatives of the operator $O(A)$ are then defined weakly by
the relation
\eq
\bra{\Psi}{\delta O(A)\over\delta A^\mu({\vec x},t)}\ket{\Psi} =
{\delta\over\delta A^\mu({\vec x},t)}\bra{\Psi} O(A)\ket{\Psi} ,
\label{eq:functional}
\ee
for all $\ket{\Psi}$ belonging to the domain of $O(A)$. The operators
$A^\nu$ themselves can be considered as functionals of the functions 
$A^\mu({\vec x},t)$, i.e.
\eq
\bra{\Psi}A^\nu\ket{\Psi} = \int{\rm d}{\vec x}
\int{\rm d}{\vec y}\, \langle\Psi\ket{\vec y}\,
A^\nu({\vec x},t)\,\delta({\vec x}-{\vec y})\langle{\vec y}\ket{\Psi}.
\ee
Then, from the definition (\ref{eq:functional}) one has
\eq
{\delta A^\nu\over\delta A^\mu({\vec x},t)} = \delta_{\mu\nu}\, \rho({\vec x}) ,
\ee
where $\rho({\vec x})$ is the density operator defined by
\eq
\bra{\vec y} \rho({\vec x})\ket{\Psi} = \delta({\vec x}-{\vec y})
\langle{\vec y}\ket{\Psi} .
\label{eq:density}
\ee

The density operator is also useful to express the operators $\lambda(t)$ and
$\dot{\lambda}(t)$. Using eq. (\ref{eq:density}) one readily obtains
\eq
\lambda(t) = \int{\rm d}{\vec x}\, \rho({\vec x})\,\lambda({\vec x},t),
\qquad
\dot{\lambda}(t) = \int{\rm d}{\vec x}\, \rho({\vec x})\,
\partial_t\lambda({\vec x},t) .
\ee
Therefore, $\lambda(t)$ and $\dot{\lambda}(t)$ can be considered as
functionals of $\lambda({\vec x},t)$ and $\partial_t\lambda({\vec x},t)$,
respectively, and
\eq
{\delta\lambda(t)\over\delta\lambda({\vec x},t)} =
{\delta\dot{\lambda}(t)\over\delta\partial_t\lambda({\vec x},t)} =
\rho({\vec x}) .
\label{eq:funclambda}
\ee

As a consequence, both sides of eq. (\ref{eq:lambda}) are functionals of 
$\lambda({\vec x},t)$ and $\partial_t\lambda({\vec x},t)$. Using eq.
(\ref{eq:funclambda}) and the law for the derivative of a composed functional,
the corresponding functional derivatives (evaluated at $\lambda({\vec x},t)=0$
and $\partial_t\lambda({\vec x},t)=0$) give (see appendix)
\eq
{\partial\over\partial x_i}
\left({\delta H(A)\over\delta A^i({\vec x},t)}\right) 
= -{\rm i} e \left[ H(A),\rho({\vec x})\right] ,
\label{eq:dodici}
\ee
\eq
{\delta  H(A)\over\delta A^0({\vec x},t)} = e\,\rho({\vec x}).
\label{eq:tredici}
\ee

Thus, the quantities
\eq
J_\mu(A,{\vec x}) \equiv {\delta H(A)\over\delta A^\mu({\vec x},t)},
\ee
satisfy the continuity equation
\eq
{\partial\over\partial x_i} J_i(A,{\vec x}) 
= -{\rm i}\left[ H(A), J_0(A,{\vec x}) \right]
\ee
and are interpreted as the components of the charge-current density operator.
From eqs. (\ref{eq:dodici}) and (\ref{eq:tredici}) it appears that 
$J_i(A,{\vec x})$ depend on $t$ only through their functional dependence
on $A^\mu$ and $J_0(A,{\vec x})$ does not depend on $A^\mu$.

Let us expand $H(A)$ into the functional Taylor series
\eq
H(A) = H_0 + \int{\rm d}{\vec x}\, j_\mu({\vec x}) A^\mu({\vec x}, t) +
\oneh\int{\rm d}{\vec x}\int{\rm d}{\vec y}\, A^\mu({\vec x}, t) 
b_{\mu\nu}({\vec x},{\vec y}) A^\nu({\vec y}, t) + \dots ,
\label{eq:expand}
\ee
where
\eq
j_\mu({\vec x}) = 
\left.{\delta H(A)\over\delta A^\mu({\vec x}, t)}\right\vert_{A=0},
\qquad
b_{\mu\nu}({\vec x},{\vec y}) =
\left.{\delta^2 H(A)\over\delta A^\mu({\vec x},t)\,\delta A^\nu({\vec y},t)}
\right\vert_{A=0} .
\ee
These operators do not depend on $t$ (see eqs. (\ref{eq:contchar}) and
(\ref{eq:contcurr}) below). In addition, due to the independence of the order
of functional differentiation the following symmetry property holds
\eq
b_{\mu\nu}({\vec x},{\vec y}) = b_{\nu\mu}({\vec y},{\vec x}) .
\ee

Now let us relate $j_\mu({\vec x})$ and $b_{\mu\nu}({\vec x},{\vec y})$ to 
$J_\mu(A,{\vec x})$. Since $H(A)$ linearly depends on $A^0$, one has
\eq
j_0({\vec x}) = J_0(A,{\vec x}) = e \rho({\vec x}),
\label{eq:charge}
\ee
\eq
b_{0\nu}({\vec x},{\vec y}) = b_{\mu 0}({\vec x},{\vec y}) = 0.
\ee
Using eq. (\ref{eq:expand}) one has
\eq
J_i(A,{\vec x}) = j_i({\vec x}) + \int{\rm d}{\vec y}\, b_{ij}({\vec x},{\vec
y}) A^j({\vec y},t) +\dots .
\label{eq:capj}
\ee
Thus, $j_i({\vec x})$ is the first-order expression of the current-density
operator. 

Inserting eqs. (\ref{eq:expand}) and (\ref{eq:capj}) into eq. (\ref{eq:dodici}),
which identically holds in $A^\mu$, one obtains
\eq
{\partial\over\partial x_i} j_i({\vec x}) = -{\rm i} e [H_0,\rho({\vec x})],
\label{eq:contchar}
\ee
\eq
{\partial\over\partial x_i} b_{ij}({\vec x},{\vec y}) = - {\rm i} e
[j_j({\vec y}),\rho({\vec x})] .
\label{eq:contcurr}
\ee

\end{section}

\begin{section}{Electromagnetic interaction in semi-rela\-tiv\-is\-tic theories}

Ignoring spin degrees of freedom for the time being, let us consider the
following semi-relativistic Hamiltonian
\eq
H_0 = \plusqrt{{\vec p}^2 + m^2} + V,
\label{eq:semirelh}
\ee
where the self-adjoint operator $V$ is energy and momentum independent. Under
the action of the electromagnetic field
\eq
A^\mu(x) = \epsilon^\mu \,{\rm e}^{{\rm i}({\vec q}\cdot{\vec x} - q_0 t)},
\ee
the minimal substitution gives
\eq
H(A) = \plusqrt{({\vec p} - e {\vec A})^2 + m^2} + e A_0 + V.
\label{eq:srham}
\ee
Since the operators $A^\mu$ are bounded, the operator $({\vec p} - e {\vec
A})^2$ is self-adjoint. Hence, also the positive square root in eq.
(\ref{eq:srham}) is a well-defined self-adjoint operator.

Due to eq. (\ref{eq:expand}), the interaction Hamiltonian with the
electromagnetic field is written (up to second order) as
\eq
H_{\rm em}(A) = H^{(1)}_{\rm em}(A) + H^{(2)}_{\rm em}(A) ,
\ee
where
\eq
H^{(1)}_{\rm em}(A) = \int{\rm d}{\vec x}\,j_\mu({\vec x}) A^\mu({\vec x},t) ,
\ee
\eq
H^{(2)}_{\rm em}(A) = \oneh\int{\rm d}{\vec x}\int{\rm d}{\vec y}\,
A^i({\vec x},t) b_{ij}({\vec x},{\vec y})A^j({\vec y},t).
\ee

\begin{subsection}{Expression of $H^{(1)}_{\rm em}(A)$}

The charge density operator $j_0$ is already known from eq. (\ref{eq:charge}).
The first-order current density operator can be written as
\eq
j_i({\vec x}) = 
\left.{\delta R(A)\over\delta A^i({\vec x},t)}\right\vert_{A=0},
\label{eq:firstcurr}
\ee
where
\eq
R(A) = \plusqrt{({\vec p} - e {\vec A})^2 + m^2} .
\ee
The r.h.s. of eq. (\ref{eq:firstcurr}) can be obtained from the easily
calculable functional derivative of $[R(A)]^2$. One has
\eq
\left\{ R(A), {\delta R(A)\over\delta A^i({\vec x},t)} \right\}_{A=0} =
\left. {\delta [R(A)]^2\over\delta A^i({\vec x},t)}\right\vert_{A=0} =
e\left\{ p_i,\rho({\vec x})\right\} .
\label{eq:anti}
\ee
The anticommutators in eq. (\ref{eq:anti}) can be solved in momentum
representation, where from eq. (\ref{eq:density}) one has
\eq
\bra{{\vec k}'}\rho({\vec x})\ket{{\vec k}} =
(2\pi)^{-3}\, {\rm e}^{{\rm i}({\vec k}-{\vec k}')\cdot{\vec x}}.
\ee
Thus eqs. (\ref{eq:firstcurr}) and (\ref{eq:anti}) give
\eq
\bra{{\vec k}'}j_\mu({\vec x})\ket{{\vec k}} = e {k_\mu + k'_\mu\over k_0 +
k'_0} (2\pi)^{-3}\, {\rm e}^{{\rm i}({\vec k}-{\vec k}')\cdot{\vec x}},
\label{eq:fourcurrent}
\ee
where
\eq
k_0 \equiv E_{\vec k} = \plusqrt{{\vec k}^2 + m^2} , \qquad
k'_0 \equiv E_{{\vec k}'} = \plusqrt{{\vec k}'{}^2 + m^2}.
\ee
One easily checks that the continuity equation (\ref{eq:contchar}) is satisfied.

From eqs. (\ref{eq:expand}) and (\ref{eq:fourcurrent}) the interaction with
the electromagnetic field is (up to first order in $e$)
\eq
\bra{{\vec k}'}H^{(1)}_{\rm em}(A)\ket{{\vec k}} = e
{\epsilon^\mu(k_\mu + k'_\mu)\over k_0 + k'_0}\,\delta({\vec k}+{\vec q} -
{\vec k}')\,{\rm e}^{-{\rm i} q_0 t}.
\label{eq:emfirst}
\ee
In terms of operators eq. (\ref{eq:emfirst}) reads
\eq
H^{(1)}_{\rm em}(A) 
= e\left(\epsilon_0 - {{\vec\epsilon}\cdot(2{\vec p} - {\vec q})\over
\plusqrt{{\vec p}^2 + m^2} + \plusqrt{({\vec p}-{\vec q})^2 + m^2} }\right)
{\tilde\rho}({\vec q}) \,{\rm e}^{-{\rm i} q_0 t},
\label{eq:accauno}
\ee
where
\eq
{\tilde\rho}({\vec q}) = \int{\rm d}{\vec x}\, \rho({\vec x})\,{\rm e}^{{\rm
i}{\vec q}\cdot{\vec x}}.
\ee

\end{subsection}

\begin{subsection}{Expression of $H^{(2)}_{\rm em}(A)$}

For the operators $b_{ij}({\vec x},{\vec y})$ the following relation
\eq
b_{ij}({\vec x},{\vec y}) = \left.{\delta^2 R(A)\over\delta A^i({\vec x},t)\,
\delta A^j({\vec y},t) }\right\vert_{A=0}
\ee
can be derived from 
\eq
\left\{ 
R(A), { \delta^2 R(A)\over\delta A^i({\vec x},t)\,\delta A^j({\vec y},t) }
\right\} = 
{ \delta^2 [R(A)]^2\over \delta A^i({\vec x},t)\,\delta A^j({\vec y},t) }
- \left\{ {\delta R(A)\over\delta A^i({\vec x},t)} , 
{\delta R(A)\over\delta A^j({\vec y},t)} \right\} .
\ee
Using eq. (\ref{eq:firstcurr}) and calculating the second-order derivative of
$[R(A)]^2$ one has
\eq
\left\{ 
R(A), { \delta^2 R(A)\over\delta A^i({\vec x},t)\,\delta A^j({\vec y},t) }
\right\} = 2 e^2\, \delta_{ij}\,\rho({\vec x})\,\rho({\vec y}) -
\left\{ j_i({\vec x}), j_j({\vec y})\right\},
\ee
so that
\eq
\bra{{\vec k}'} b_{ij}({\vec x},{\vec y}) \ket{{\vec k}} = 
{1\over k_0 + k'_0} \bra{{\vec k}'}\left[ 2\,\delta_{ij}\, 
j_0({\vec x})\,j_0({\vec y}) - \left\{ j_i({\vec x}), j_j({\vec y})\right\}
\right]\ket{{\vec k}} .
\ee
Hence, the second-order correction of the electromagentic interaction is
\eqa
\bra{{\vec k}'} H^{(2)}_{\rm em}(A) \ket{{\vec k}} &=&
{e^2\over E_{\vec k} + E_{{\vec k}'} }
\left[ {\vec\epsilon}\cdot{\vec\epsilon}
+ {({\vec\epsilon}\cdot{\vec q})^2 - 4( {\vec\epsilon}\cdot
({{\vec k}'} - {\vec q}) )^2 \over 
(E_{{\vec k}'} + E_{{\vec k}' -{\vec q}})
(E_{\vec k} + E_{{\vec k}' -{\vec q}}) }
\right]\nonumber \\
& & \nonumber \\
&{}&\quad\qquad\times\,
\delta({\vec k} + 2{\vec q} - {\vec k}') \,{\rm e}^{-2{\rm i}q_0 t} .
\label{eq:accadue}\\ \nonumber
\eea

\end{subsection}

\begin{subsection}{Spin contribution}

To treat a particle with spin $\oneh$, one endows the wavefunctions with the
matrix structure related to the Pauli matrices $\sigma_i$ and writes the
Hamiltonian $H_0$ of eq. (\ref{eq:semirelh}) in the equivalent form
\eq
H_0 = \plusqrt{({\vec\sigma}\cdot{\vec p})^2 + m^2} + V.
\label{eq:semirelhspin}
\ee
Therefore, the minimal substitution gives
\eq
H(A) = \plusqrt{({\vec p} - e{\vec A})^2 - e{\vec\sigma}\cdot{\vec B} + m^2} 
+ e A_0 + V,
\ee
where
\eq
{\vec B} = {\vec\nabla}\times{\vec A}.
\ee
The first-order contribution of the spin to the charge-current density,
\eq
j_\mu^S({\vec x}) = \left.
{\delta H^S(A)\over\delta A^\mu({\vec x},t)}\right\vert_{A=0},
\ee
where
\eq
H^S(A) = H(A) - \plusqrt{({\vec p} - e{\vec A})^2  + m^2} - e A_0 - V,
\ee
can be calculated applying the method of sect. 3.1 to each single term in 
$H^S(A)$ and using the relations
\eq
{\delta{\vec\sigma}\cdot{\vec B}\over\delta A^i({\vec x},t) } =
{\rm i} {\delta\over\delta A^i({\vec x},t) }\left[
({\vec\sigma}\times{\vec p})\cdot{\vec A} - 
 {\vec A}\cdot({\vec\sigma}\times{\vec p})\right] =
-{\rm i}\left[({\vec\sigma}\times{\vec p})_i,\rho({\vec x})\right] ,
\ee
\eq
{\delta{\vec\sigma}\cdot{\vec B}\over\delta A^0({\vec x},t)} = 0.
\ee
One obtains
\eq
\bra{{\vec k}',s'} {\vec j}^S({\vec x})\ket{{\vec k},s} = {{\rm i} e\over
k_0 + k'_0} \bra{s'}{\vec\sigma}\times({\vec k}'-{\vec k})\ket{s}
(2\pi)^{-3}\,
e^{{\rm i}({\vec k}-{\vec k}')\cdot{\vec x}},
\ee
\eq
\bra{{\vec k}',s'} {\vec j}_0({\vec x})\ket{{\vec k},s} = 0.
\ee
The corresponding contribution to the first-order interaction Hamiltonian is
\eq
\bra{{\vec k}',s'} H^{S(1)}_{\rm em}(A)\ket{{\vec k},s} = -
{e\over k_0 + k'_0}
 \bra{{\vec k}',s'} {\vec\sigma}\cdot{\vec B}\ket{{\vec k},s} .
\ee

\end{subsection}

\end{section}

\begin{section}{Application to quasi-elastic electron scattering}

In order to compare the semi-relativistic charge-current density operator
obtained in sect. 3 with other operators existing in the literature we
consider quasi-elastic electron scattering. The advantage of this process is
that it is almost independent of any details of the nuclear structure. In
fact, neglecting final-state interactions the coincidence (e,e$'$p)
cross section for unpolarized electrons is simply proportional to the
(off-shell) electron-proton cross section $\sigma_{\rm ep}$~\cite{[book]},
\eq
{{\rm d}^3\sigma\over {\rm d}E'\, {\rm d}\Omega \, 
{\rm d}\Omega _{\rm p}} = K \sigma_{\rm ep}\, S({\vec p}, E) ,
\ee
where $K$ is a kinematical factor and $S({\vec p},E)$ is the spectral function
for knocking out a proton of momentum ${\vec p}$ and energy $E$. For a
coplanar kinematics,
\eq
\sigma_{\rm ep}=
\sigma_{\rm M} (\rho_{00}\, g_{00} + \rho_{11}\, 
g_{11} +  \rho_{01}\, g_{01} + \rho_{1-1}\, g_{1-1}),
\ee
where $\sigma_{\rm M}$ is the Mott cross section describing elastic Coulomb 
scattering by a point-like particle. The coefficients
$\rho_{\lambda\lambda'}$ only depend on the electron kinematic variables and
are well known from quantum electrodynamics~\cite{[book]}. Thus, the four
structure functions $g_{\lambda\lambda'}$ (with $\lambda = 0$ for the
longitudinal  polarization and $\lambda = \pm 1$ for the two transverse
polarizations of the exchanged photon) are the relevant quantities to study the
effects due to the different charge-current density operators. In particular,
$g_{00}$ and $g_{11}$ contain the pure contribution of the charge and current
density, respectively, while interference terms produce $g_{01}$ and $g_{1-1}$.

The $\sigma_{\rm ep}$ most commonly used was proposed by de
Forest~\cite{[deForest]}, who made a detailed comparison of the existing cross
sections in different kinematical situations. The nuclear current of eq.
(\ref{eq:cc1}) produces the cross section $\sigma_{\rm cc1}$, while the one of
eq. (\ref{eq:cc2}) gives $\sigma_{\rm cc2}$. On shell the two cross sections
are equal, but even off shell they are very close to each other. However,
$\sigma_{\rm cc1}$, which is widely used, is simpler and using on-shell
kinematics the separated contributions to $\sigma_{\rm ep}$ are given by

\eqa
g_{00} 
&=& {(E_{\vec p} +  E_{{\vec p}'})^2\over 4 E_{\vec p} E_{{\vec p}'}} 
\left (F_1^2 +  {Q^2 \over 4 M^2} \kappa^2 F^2_2  \right )
 - {1\over 4 E_{\vec p} E_{{\vec p}'}} 
\vert{\vec q}\vert^2 (F_1 +  \kappa F_2)^2, \nonumber \\
& & \nonumber \\
g_{11} 
&=& {\vert {\vec p}'\vert^2 \sin^2 \gamma \over E_{\vec p} 
E_{{\vec p}'}} \left (F_1^2 +  {Q^2 \over 4 M^2} \kappa^2
F^2_2  \right ) +
{Q^2 \over 2 E_{\vec p} E_{{\vec p}'}} (F_1 +   \kappa
F_2)^2 , \nonumber \\
& & \nonumber \\ 
g_{01} 
&=& - {{\sqrt 2} (E_{\vec p} + E_{{\vec p}'})\over E_{\vec p} E_{{\vec p}'}} 
\vert {\vec p}'\vert \sin \gamma \left (F_1^2 + 
{Q^2 \over 4 M^2} \kappa^2 F^2_2 \right ), \nonumber \\ 
& &\nonumber \\ 
g_{1-1} 
&=& -{\vert {\vec p}'\vert^2 \sin^2 \gamma \over E_{\vec p} 
E_{{\vec p}'}} \left (F_1^2 +  {Q^2 \over 4 M^2} 
\kappa^2 F^2_2 \right ). 
\label{eq:sigmacc1}\\ \nonumber
\eea
where $\gamma$ is the angle between ${\vec q}$ and ${\vec p}'$,
\eq
E_{\vec p} = \sqrt{\vert {\vec p}\vert^2 +  M^2}, \qquad
E_{{\vec p}'} = \sqrt{\vert {\vec p}'\vert^2 +  M^2}
\ee
are the initial and final proton energy, respectively, and
\eq
\omega = E_{{\vec p}'} - E_{\vec p}
\ee
is the energy transfer with ${\vec p}' = {\vec p} + {\vec q}$.

The same structure functions calculated with the non-relativistic nucleon
charge-current operator and truncated to order $M^{-2}$ are given by
\eqa
g_{00} &=& F_1^2 - F_1(F_1 + 2\kappa F_2){1\over 4M^2}Q^2 ,\nonumber \\
& & \nonumber \\
g_{11} &=& F_1^2{1\over M^2}\vert {\vec p}'\vert^2\sin^2\gamma 
+ (F_1 + \kappa F_2)^2{1\over 2M^2}\vert{\vec q}\vert^2,\nonumber \\
& & \nonumber \\
g_{01} 
&=& -2\sqrt{2} F_1^2{1\over M}\vert {\vec p}'\vert\sin\gamma,\nonumber \\
& & \nonumber \\
g_{1-1} &=& - F_1^2{1\over M^2}\vert {\vec p}'\vert^2\sin^2\gamma  .
\label{eq:nonrel}\\ \nonumber
\eea
Higher-order corrections are obtained following the approach of
ref.~\cite{[GP]}.

The corresponding expressions in the semi-relativistic case are:
\eqa
g_{00} &=& F_1^2,\nonumber \\
& & \nonumber \\
g_{11} &=& F_1^2 {4\over(E_{\vec p}+ E_{{\vec p}'})^2}
\vert {\vec p}'\vert^2\sin^2\gamma  + (F_1 + \kappa F_2)^2
{2\over(E_{\vec p}+ E_{{\vec p}'})^2}\vert{\vec q}\vert^2,\nonumber \\
& & \nonumber \\
g_{01} &=& -2\sqrt{2} F_1^2 {2\over E_{\vec p}+ E_{{\vec p}'}}
 \vert {\vec p}'\vert \sin\gamma,\nonumber \\
& & \nonumber \\ 
g_{1-1} &=& - F_1^2 {4\over(E_{\vec p}+ E_{{\vec p}'})^2}
\vert {\vec p}'\vert^2 \sin^2\gamma  .
\label{eq:semirel}\\ \nonumber
\eea
In the semi-relativistic case the charge and current densities include form
factors in order to describe an extended particle and, correspondingly, the
contribution of the anomalous magnetic moment has been added to the spin
current.

There are two differences between the semi-relativistic and non-relativistic
structure functions. First, the Darwin-Foldy (DF) term appearing in the
non-relativistic $g_{00}$ cannot be reproduced in the semi-relativistic case,
because only the spatial components of the charge-current density operator are
affected when going from the non-relativistic kinetic energy to the
semi-relativistic one in eq. (\ref{eq:semirelh}), while the time component is
simply given by the charge density. In fact, the DF term arises in the
non-relativistic reduction of the Dirac equation which has negative-energy
solutions, whereas the semi-relativistic Hamiltonian does not have such
solutions. Second, apart from the DF term, the semi-relativistic structure
functions are obtained from the non-relativistic ones by substituting the mass
$M$ with $\oneh(E_{\vec p} + E_{{\vec p}'})$ so that they coincide in the
low-energy limit, $E\approx M$. In this limit they also become equal to the
structure functions of the relativistic $\sigma_{\rm cc1}$ apart from the
$Q^2$-dependent terms involving the anomalous magnetic moment in the convective
current and the charge density. Precisely these terms are responsible for the
observed deviations from a non-relativistic description of the quasi-free
knockout in the standard kinematics explored up to now. Of course,
larger deviations between the relativistic and non-relativistic structure
functions are expected for increasing energy and momentum transfers.

In order to study the behaviour of the current given by the different
approaches one has to focus on the pure transverse structure function,
$g_{11}$. Assuming the result $g_{11}({\rm cc1})$ obtained within the cross
section $\sigma_{\rm cc1}$ as a reference, the percentage deviation from this
result calculated within the other approaches,
\eq
\Delta = {g_{11} - g_{11}({\rm cc1})\over g_{11}({\rm cc1})}\times 100,
\ee
will be considered as an indication of how far one is from the correct
relativistic result.

In fig. 1 results are shown in parallel kinematics, i.e. when ${\vec q}$
and ${\vec p}'$ are aligned, for three values of $p'= 0.5$, 0.75 and 1 GeV.
The socalled perpendicular kinematics is used in figs. 2 and 3, i.e. $q$ and
$p'$ are fixed and the angle $\gamma$ between ${\vec q}$ and ${\vec p}'$ is
varying. In fig. 2, $p'=0.5$ GeV, while in fig. 3, $p'=1$ GeV. The corresponding
results for $p'=0.75$ GeV can be interpolated from those in figs. 2 and 3. For
$\gamma = 0^\circ$ one recovers the result in parallel kinematics at the
corresponding $q$ value.

As already known from the analysis of ref.~\cite{[GP]}, the non-relativistic
result (to order $M^{-2}$) and that corrected to order $M^{-3}$ are always of
opposite sign with respect to the result corrected to order $M^{-4}$. The
latter is within a few percent from the relativistic case for $p'=0.5$ GeV
both in parallel and perpendicular kinematics. However, for larger momenta the
deviation rapidly deteriorates, in agreement with ref.~\cite{[McVoy]}, where
it was stressed that the non-relativistic FW reduction becomes meaningless for
values of $p/M$ larger than 0.5, with $p$ a typical momentum. On the contrary,
the semi-relativistic result is comparable with the non-relativistic one at
low energy and soon merges into the relativistic one for increasing $q$ and
$p'$. 

\end{section}

\begin{section}{Concluding remarks}

Starting from a single-particle semi-relativistic Hamiltonian, where only the
kinetic part has a relativistic shape, the corresponding charge-current
density and two-photon operators have been obtained within a suitable
functional derivative formalism. Such operators satisfy gauge invariance by
construction and fully describe the behaviour of a point particle undergoing
an external electromagnetic interaction as long as the effects of
negative-energy states can be neglected. 

In particular, the matrix elements of the current density operator between
free-particle states coincide with the  non-relativistic expressions of the
usual convection and spin currents only in the low-energy limit when the total
energy of the particle can simply be approximated by its mass. On the other
hand, once eliminated the presence of negative-energy states {\sl ab initio}, no
corrections to the non-relativistic charge operator are possible in the present
semi-relativistic approach. Nonetheless, the operators obtained here are valid
for energies up to the threshold of particle-antiparticle production, thus
substantially increasing the range of applicability with respect to any
non-relativistic reduction of the electromagnetic interaction. In fact, it is
well known that non-relativistic charge-current density operators, even
corrected up to fourth order in $M^{-1}$, do not apply for values of $p/M$
larger than 0.5, with $p$ a typical momentum. On the contrary, a comparison
between results obtained with the semi-relativistic current density operator
and a fully relativistic operator presented in the case of electron-proton
scattering shows a more and more satisfactory agreement for increasing values of
the ejectile momentum and/or momentum transfer. 

Therefore, whenever the impulse approximation applies, the proposed 
semi-relativistic operators could also be used to describe the electromagnetic
interaction of a many-body system at intermediate energies.

\end{section}

\bigskip

This work has been performed in part under the contract ERB FMRX-CT-96-0008
within the frame of the Training and Mobility of Researchers Programme of the
Commission of the European Union.

\clearpage

\centerline{\Large{\bf Appendix}}

\bigskip

In this appendix the derivation of eqs. (\ref{eq:dodici}) and 
(\ref{eq:tredici}) is given.

$H(A')$ is a functional of the functions ${A'}^i({\vec y},t) 
= A^i({\vec y},t) - \partial\lambda({\vec y},t)/\partial y_i$ and 
${A'}^0({\vec y},t) 
= A^0({\vec y},t) - \partial\lambda({\vec y},t)/\partial t$ which in turn are
functionals (dependent on the parameter ${\vec y}$) of the functions
$\lambda({\vec x},t)$ and $\partial\lambda({\vec x},t)/\partial t$:
\eq
{A'}^i({\vec y},t) = A^i({\vec y},t) +
\int{\rm d}{\vec x}\,{\partial\delta({\vec x} - {\vec y})\over\partial x_i}
\lambda({\vec x},t) ,
\label{eq:appone}
\ee
\eq
{A'}^0({\vec y},t) = A^0({\vec y},t) -
\int{\rm d}{\vec x}\,\delta({\vec x} - {\vec y})
{\partial\lambda({\vec x},t)\over\partial t} .
\label{eq:apptwo}
\ee
By the law of differentiation of a composed functional one obtains
\eq
{\delta H(A')\over\delta\lambda({\vec x},t)} =
\int{\rm d}{\vec y}\,{\delta H(A')\over\delta {A'}^i({\vec y},t)}
{\delta {A'}^i({\vec y},t)\over\delta\lambda({\vec x},t)} ,
\ee
\eq
{\delta H(A')\over\delta(\partial_t \lambda({\vec x},t))} =
\int{\rm d}{\vec y}\,{\delta H(A')\over\delta {A'}^0({\vec y},t)}
{\delta {A'}^0({\vec y},t)\over\delta(\partial_t \lambda({\vec x},t))},
\ee
where, due eqs. (\ref{eq:appone}) and (\ref{eq:apptwo}),
\eq
{\delta {A'}^i({\vec y},t)\over\delta\lambda({\vec x},t)} =
 {\partial\delta({\vec x} - {\vec y})\over\partial x_i},
\ee
\eq
{\delta {A'}^0({\vec y},t)\over\delta(\partial_t \lambda({\vec x},t))} =
- \delta({\vec x} - {\vec y}) .
\ee
Therefore,
\eq
\left. {\delta H(A')\over\delta\lambda({\vec x},t)}\right\vert_{\lambda=0}
=  \int{\rm d}{\vec y}\,{\delta H(A)\over\delta {A}^i({\vec y},t)}
{\partial\delta({\vec x} - {\vec y})\over\partial x_i} = 
{\partial\over\partial x_i} {\delta H(A)\over\delta {A}^i({\vec x},t)} ,
\label{eq:appseven}
\ee
\eq
\left. {\delta H(A')\over\delta(\partial_t \lambda({\vec x},t))}
\right\vert_{\lambda=0} = - \int{\rm d}{\vec y}\,
{\delta H(A)\over\delta {A}^0({\vec y},t)} \delta({\vec x} - {\vec y}) 
= - {\delta H(A)\over\delta {A}^0({\vec y},t)} .
\label{eq:appeight}
\ee

One is now left with the evaluation of the functional derivative of 
${\rm e}^{{\rm i}e\lambda(t)}$ which is a function of the functional
$\lambda(t)$ of $\lambda({\vec x},t)$. Due to the first eq.
(\ref{eq:funclambda}) one has:
\eq
{\delta\, {\rm e}^{{\rm i}e\lambda(t)}\over\delta\lambda({\vec x},t)} =
{\rm i} e \,{\rm e}^{{\rm i}e\lambda(t)}\,\rho({\vec x}) 
\Longrightarrow \left.
{\delta\, {\rm e}^{{\rm i}e\lambda(t)}\over\delta\lambda({\vec x},t)}
\right\vert_{\lambda=0} = {\rm i} e \,\rho({\vec x}).
\label{eq:appnine}
\ee
Using eqs. (\ref{eq:appseven}), (\ref{eq:appeight}), (\ref{eq:appnine}) and the
second eq.  (\ref{eq:funclambda}) one obtains eqs. (\ref{eq:dodici}) and 
(\ref{eq:tredici}).

\clearpage

%%%%%%%%%%%%%%%%%%%%%%%%%%%%%  References  %%%%%%%%%%%%%%%%%%%%%%%%%%%%%%%%

\clearpage

\centerline{\Large{\bf Figure captions}}

\bigskip

\noindent
Fig. 1. The percentage deviation $\Delta$ of the pure transverse structure
function $g_{11}$ from the value obtained within the cross section $\sigma_{\rm
cc1}$, calculated within the semi-relativistic approach and the
non-relativistic ones corrected through second ($1/m^2$), third ($1/m^3$) and
fourth order ($1/m^4$). At the indicated values of the momentum of the 
proton ejected under parallel kinematic conditions, i.e. ${\vec p}'$ parallel
to the momentum transfer ${\vec q}$, $\Delta$ is plotted as a function of $q$.

\bigskip

\noindent
Fig. 2. The percentage deviation $\Delta$ of the pure transverse structure
function $g_{11}$ from the value obtained within the cross section $\sigma_{\rm
cc1}$, calculated within the semi-relativistic approach and the
non-relativistic ones corrected through second ($1/m^2$), third ($1/m^3$) and
fourth order ($1/m^4$). At the indicated values of the momentum transfer $q$,
$\Delta$ is plotted for $p'=0.5$ GeV as a function of the angle $\gamma$
between ${\vec q}$ and the the momentum ${\vec p}'$ of the ejected proton.

\bigskip

\noindent
Fig. 3. The same as in fig. 2 for $p'=1$ GeV.

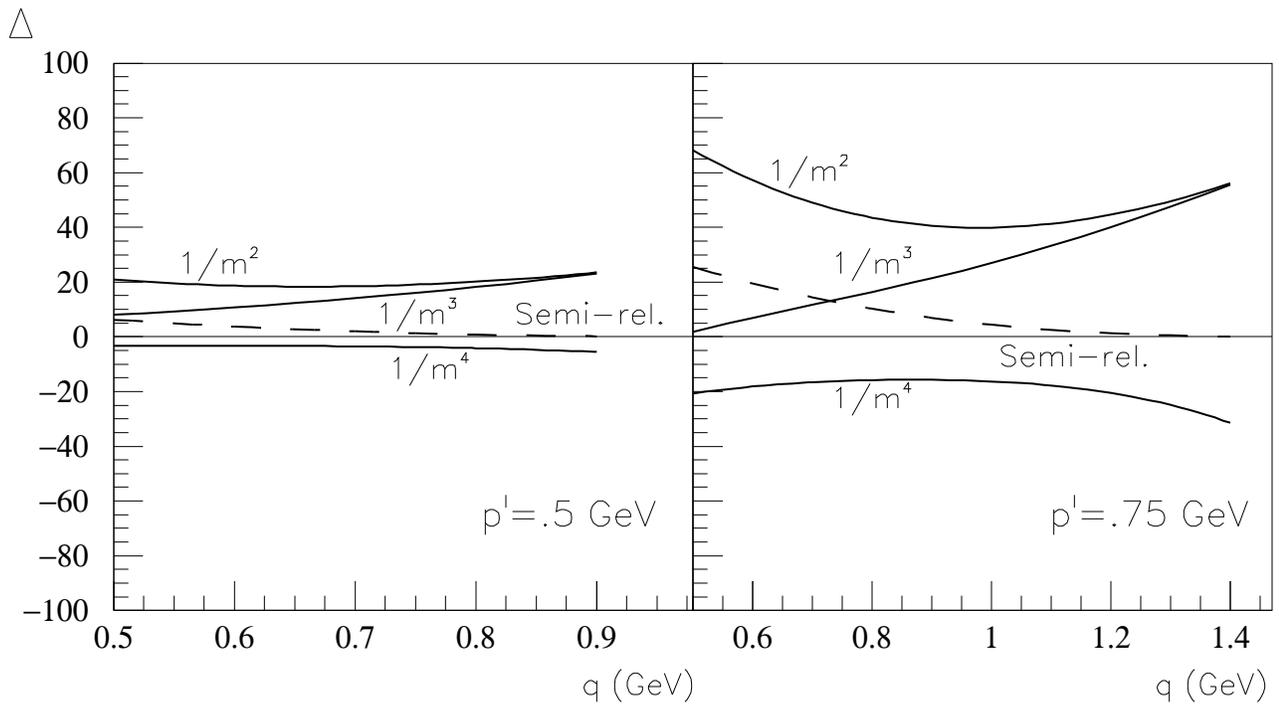
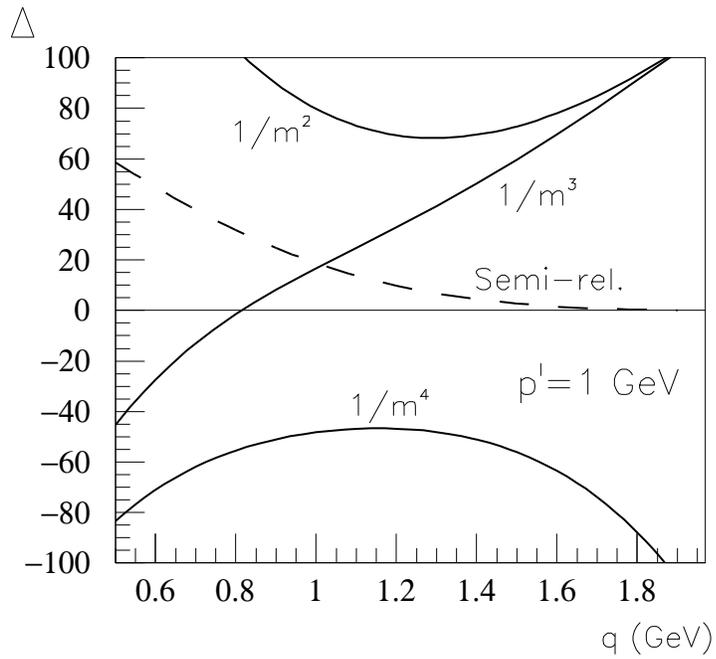
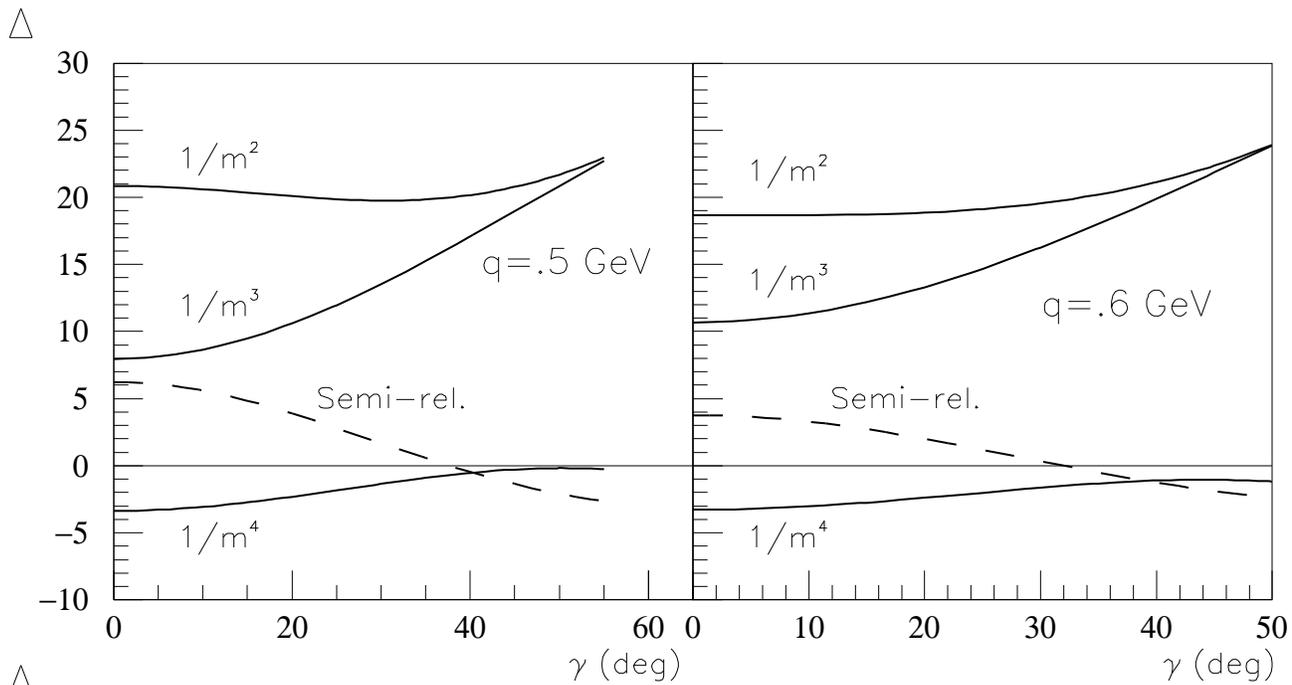
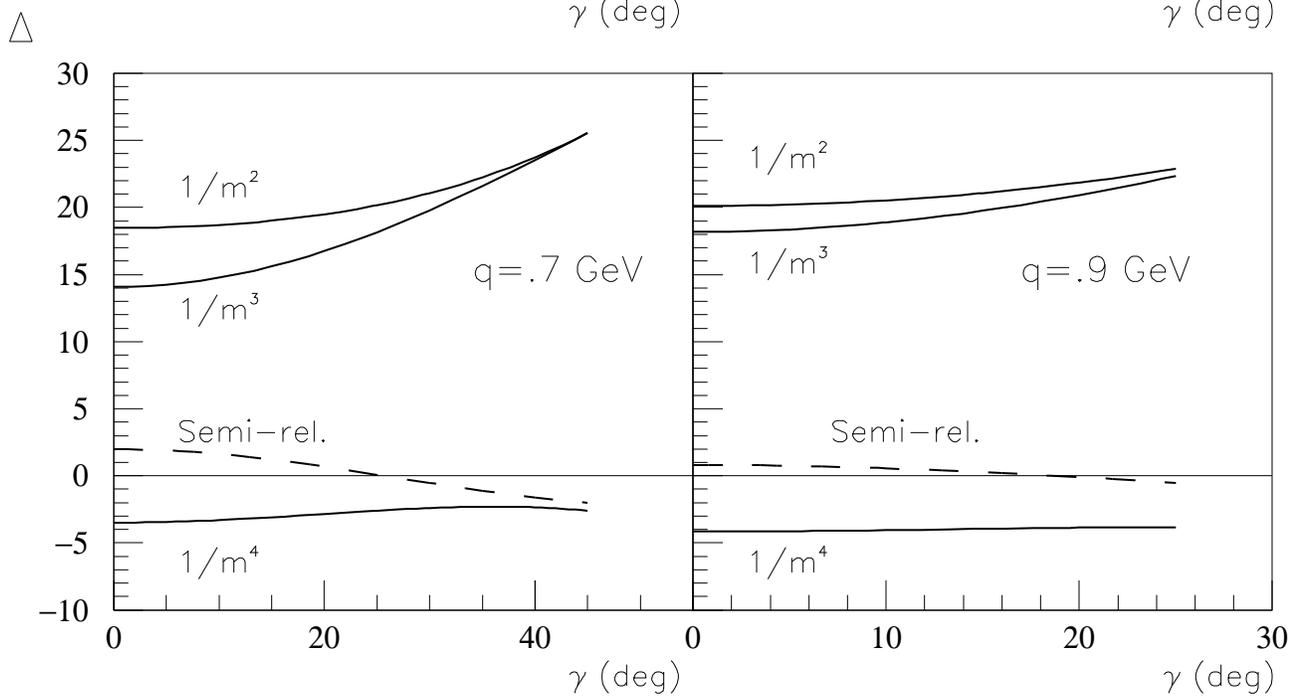
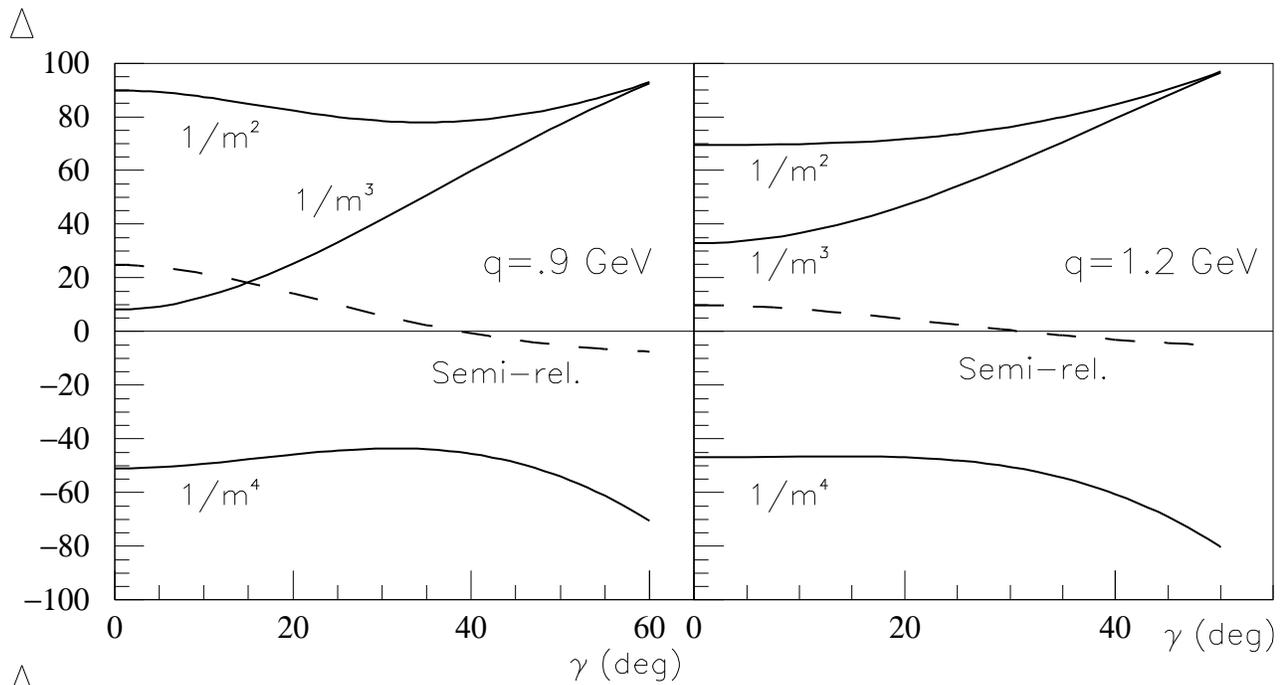
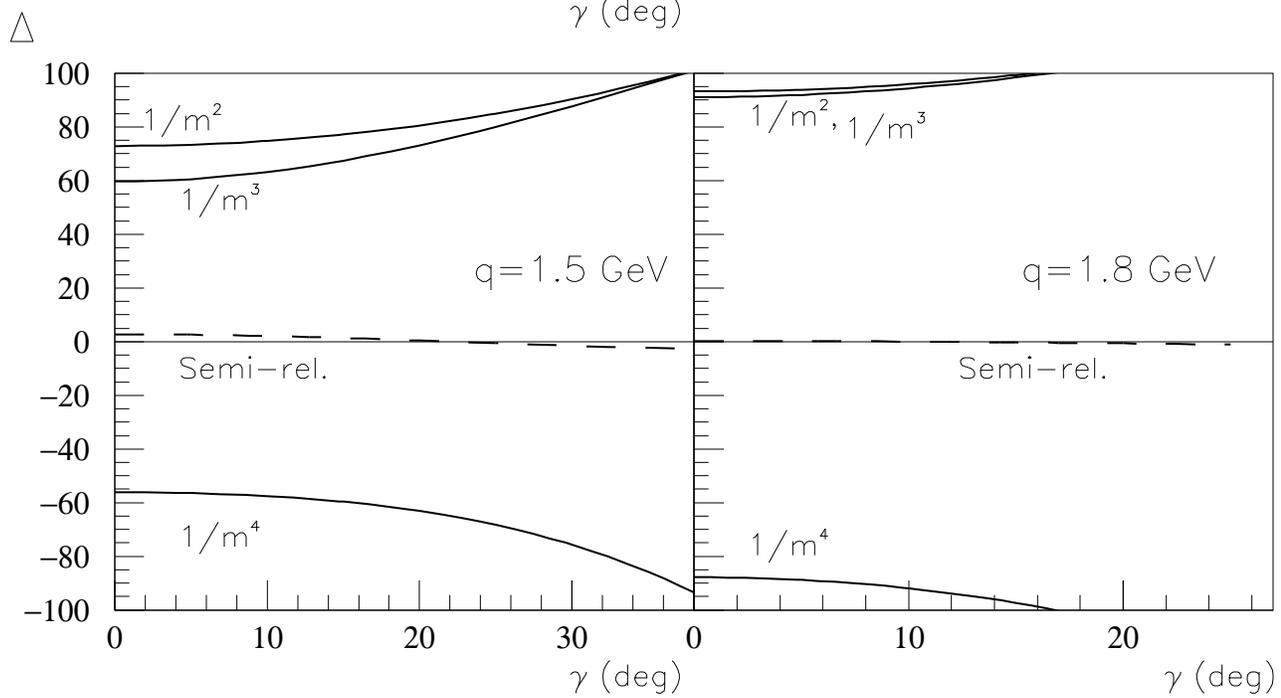
\end{document}